\documentclass[ twocolumn,aps,prd,   
               preprintnumbers,numbers,sort&compress,nofootinbib,
                            showpacs,superscriptaddress,
               colorlinks,
               linkcolor=blue,   
               citecolor=blue]{revtex4-1}   \newcommand{\exclude}[1]{}

\usepackage{graphicx,amsmath,amssymb,bm}

\usepackage{psfrag}
\usepackage{hyperref}
\usepackage{enumitem}
\usepackage{siunitx}

\newcommand{\beq}{\begin{equation}}
\newcommand{\eeq}{\end{equation}}
\newcommand{\be}{\begin{eqnarray}}
\newcommand{\ee}{\end{eqnarray}}
   
\def\dd{ \,\mathrm{d} }

\def\+{\dagger}
 \def\la{\langle}
 \def\ra{\rangle}

\begin{document}
 
\title{ A few thoughts on $\theta$ and the electric dipole moments.}

\author{Ariel \surname{Zhitnitsky}}
\email{arz@phas.ubc.ca}
\affiliation{Department of Physics and Astronomy, University of British Columbia, Vancouver, B.C. V6T 1Z1, Canada}

\begin{abstract}
 I highlight a few thoughts on  the contribution to the dipole moments    from the so-called  $\theta$ parameter. 
 The dipole moments are known  can be generated by   $\theta$. In fact, the renowned strong $\cal{CP}$ problem was formulated as a result of non-observation of the dipole moments. 
  What is less known is that  there is another parameter of the theory, the $\theta_{QED}$  which becomes  also a physical and observable parameter 
 of the system when some conditions are met. This claim should be  contrasted with conventional (and very naive) viewpoint that the $\theta_{\rm QED}$  is unphysical and unobservable.  A specific manifestation of this phenomenon is  
 the so-called  Witten effect when the magnetic monopole becomes the dyon with   induced electric charge   $e'=-e \frac{\theta_{QED}}{2\pi}$.  We argued that the similar arguments suggest that the electric magnetic dipole moment  $\mu$ of any microscopical configuration in the background of $\theta_{QED}$ generates the electric dipole moment $\la d_{\rm ind} \ra $ proportional to $\theta_{QED}$, i.e.  $\la d_{\rm ind} \ra= -   \frac{\theta_{\rm QED} \cdot  \alpha}{\pi} \mu$. 
 We also argue that many $\cal{CP}$ correlations such as $ \la \vec{B}_{\rm ext}  \cdot\vec{E}\ra  = -\frac{\alpha\theta_{\rm QED}}{\pi}\vec{B}^2_{\rm ext}$ will be generated in the background of an external   magnetic field $\vec{B}_{\rm ext}  $  as a result of the same physics. 
\end{abstract}


\maketitle

\baselineskip=15pt


\section{Introduction and Motivation}\label{introduction}
The leitmotiv of the present work is related to the fundamental parameter $\theta$ in  Quantum Chromodynamics (QCD), as well as the axion field related to this parameter. The $\theta$ parameter was originally introduced in  the 70s. Although the $\theta$ term can be represented as a total derivative and does not change the equation of motion, it is known that this parameter is a fundamental physical parameter of the system on the non-perturbative level. It is known that the $\theta\neq 0$ introduces $\cal{P}$ and $\cal{CP}$ violation in QCD, which is most well captured by the renowned strong $\cal{CP}$ problem.  

In particular, what is the most important element for the present notes is that the $\theta$ parameter generates the neutron (and proton) dipole moment which is known to be very small, $d_n\lesssim 10^{-26}{\rm e\cdot cm}$, see e.g. review in Physics Today \cite{vanBibber:2006rb}. It can be translated to the upper limit for $\theta\lesssim 10^{-10}$. The strong $\cal{CP}$ problem is formulated as follows: why parameter $\theta$ is so small in strongly coupled gauge theory?
The proton electric dipole moment $d_p$, similar to the neutron dipole moment $d_n$ will be also generated as a result of non-vanishing $\theta$. In particular, a future  measurement of the $d_p$
on the level $d_p\lesssim 10^{-29}{\rm e\cdot cm}$ will be translated to much better upper limit for $\theta\lesssim 10^{-13}$.  
 
  The strong $\cal{CP}$ problem in  QCD problem was resolved by promoting the fundamental parameter $\theta$ to a dynamical axion $\theta(x)$ field, see original papers \cite{axion1, axion2, axion3, KSVZ1, KSVZ2, DFSZ1, DFSZ2} and   review articles  \cite{Asztalos:2006kz, Rosenberg:2015kxa,Marsh:2015xka,Graham:2015ouw,Irastorza:2018dyq,Sikivie:2020zpn}. However, the axion has not yet been discovered 45 years after its initial formulation. Still, it remains the best   resolution of the  strong $\cal{CP}$ problem to date, which has also led to numerous proposals for direct dark matter searches.

On the other hand, one may also discuss a similar theta term in QED. It is normally assumed that  the $\theta_{\rm QED}$ parameter in the abelian Maxwell Electrodynamics is unphysical 
and can be always  
removed from the system. The arguments are   based on the observation that the $\theta_{\rm QED}$ term does not change the equation of motion, which is also correct for non-abelian QCD.  However, in contrast with QCD when $\pi_3[SU(3)]=\mathbb{Z}$, the  topological mapping  for the abelian gauge group $\pi_3[U(1)]=0$ is trivial. This justifies the widely accepted view that $\theta_{\rm QED}$ does not modify the equation of motions (which is correct) and does not affect any physical observables and can be safely removed from the theory (which is incorrect as we argue below).
We emphasize here that the   claim is not that  $\theta_{\rm QED}$ vanishes. Instead, the (naive) claim is that the physics cannot depend on  $\theta_{\rm QED}$ irrespective to its value.

While these arguments are indeed correct for a trivial  vacuum background when the  theory is defined on an infinitely large 3+1 dimensional Minkowski space-time, 
it has been known for quite sometime that 
the $\theta_{\rm QED}$ parameter is in fact a physical parameter of the system when the theory is formulated  on a non-simply connected, compact manifold with 
non-trivial    $\pi_1[U(1)]=\mathbb{Z}$, when the gauge cannot be uniquely fixed,  see the original references  \cite{Witten:1995gf,Verlinde:1995mz} and review \cite{Olive:2000yy}. Such a construction can be achieved, for example, by putting a system into a background of the magnetic field or defining  a system on a compact manifold with non-trivial topology.  In what follows we treat 
 $\theta_{\rm QED}$ as a new fundamental (unknown) parameter of the theory. 

 Roughly speaking, 
the phenomena, in all respects, are very similar to the Aharonov-Bohm and Aharonov Casher effects when the system is highly sensitive to pure gauge (but topologically nontrivial) configurations. In such circumstances  the system cannot be fully described by a single ground state\footnote{\label{modular}We refer to \cite{PhysRevD.40.4178} with physical explanation (in contrast with very mathematical papers mentioned above) of why the gauge cannot be uniquely  fixed in such circumstances. In paper \cite{PhysRevD.40.4178} the so-called ``modular operator" has been introduced into the theory. 
The $\exp(i\theta)$ parameter in QCD is the eigenvalue of the large gauge transformation opeartor, while  $\exp(i\theta_{\rm QED})$ is the eigenvalue of the modular operator from  \cite{PhysRevD.40.4178}. This  analogy explicitly shows why $\theta_{\rm QED}$ becomes a physically observable parameter in some circumstances.}. Instead, there are multiple  degenerate   states which are  classified by a  topological index. The physics related to pure gauge configurations describing the topological sectors  is highly nontrivial. In particular,   the gauge cannot be fixed and defined uniquely in such systems.  
  This is precisely a deep reason why    $\theta_{\rm QED}$  parameter  enters the physical observables in the axion Maxwell electrodynamics in full agreement with very generic arguments  \cite{Witten:1995gf,Verlinde:1995mz,Olive:2000yy}. Precisely these contributions   lead to the explicit $\theta_{\rm QED}$-dependent 
  effects, which cannot be formulated in terms of conventional propagating degrees of freedom (propagating photons with two physical polarizations).  

The possible physical effects from $\theta_{\rm QED}$ have also been  discussed previously \cite{Hsu:2010jm,Hsu:2011sx} in the   spirit of the present notes.
  We refer to our paper \cite{Cao:2017ocv} with explicit  and detail computations of different observable effects (such as induced dipole moment, induced current on a ring, generating the potential difference on the plates, etc) when the system is defined on a nontrivial manifold, or placed in the background of the magnetic field. 

It is important  to emphasize that some effects can be proportional to $\theta_{\rm QED}$, as opposed to ${\dot{\theta}_{\rm {QED}}}$ as commonly assumed or discussed for perturbative computations. Precisely this feature has the  important applications when  some observables are proportional to the  static time-independent  $\theta_{\rm QED} $, and, in general,  do not vanish even when $\dot{\theta}_{\rm QED}\equiv 0$, see below.

 \section{Axion $\theta$ field and variety of topological phenomena }\label{sect:axion}
Our  starting point is the demonstration that  the $\theta_{\rm QED}$ indeed does not enter the equations of motion.   As a direct consequence of this observation, the corresponding  Feynman diagrams at any perturbation order will produce vanishing result for any physical observable at constant $\theta_{\rm QED}$.
   Indeed, 
 \be
 \label{current}
 \vec{j}_a=  -\dot{\theta}_{\rm QED}~\frac{ \alpha}{2\pi} ~\vec{B}, ~~ \alpha\equiv \frac{e^2}{4\pi}, 
 \ee
 which shows that  $\dot{\theta}_{\rm QED}$ and not ${\theta}_{\rm QED}$ itself enters the equations of motion.  In our analysis we ignored spatial derivatives ${\partial_i} \theta_{\rm QED}$ as they are small for non-relativistic axions. 
This anomalous current (\ref{current}) points  along magnetic field in contrast with ordinary $E\&M$, where the current is always orthogonal to $\vec{B}$.
Most of the recent proposals \cite{Asztalos:2006kz, Rosenberg:2015kxa,Marsh:2015xka,Graham:2015ouw,Irastorza:2018dyq,Sikivie:2020zpn} to detect the dark matter   axions    are precisely based on this extra current (\ref{current})  when $\dot{\theta}$ is identified with propagating axion field oscillating with frequency $m_a$.

  We would like to make a few comments on the unusual features of this current.
  First of all, the generation of the very same non-dissipating current (\ref{current}) in the presence of $\theta$
  has been very active area of research  in recent years.  However, it is with drastically different scale of order $\Lambda_{\rm QCD}$ instead of $m_a$. The main driving force for this activity stems from the ongoing experimental results 
  at RHIC (relativistic heavy ion collider) and the LHC (Large Hadron Collider), which can be interpreted as the observation of such anomalous current (\ref{current}).
  
  The basic idea for such an interpretation can explained as follows. It has been suggested by \cite{Kharzeev:1998kz,Buckley:1999mv} that the so-called  $\theta_{\rm ind}$-domain can be formed in heavy ion collisions as a result of some non-equilibrium dynamics. This induced $\theta_{\rm ind}$ plays the same role as fundamental $\theta$ 
 and leads to a number of $\cal{P}$ and $\cal{CP}$ odd effects, such as chiral magnetic effect, chiral vortical effect, and charge separation effect, to name just a few.   
 This field  of research initiated in \cite{Kharzeev:2007tn}  became a hot topic  in recent years as a result of many interesting theoretical and experimental advances,    see recent review papers \cite{Kharzeev:2009fn,Kharzeev:2015znc} on the subject. 
 
 In particular, the charge separation effect mentioned above can be viewed as a result of generating of the induced electric field  
 \be
 \label{E_ind}
 \la \vec{E}\ra_{\rm ind} = -\frac{\alpha\theta_{\rm QED}}{\pi}\vec{B}_{\rm ext} 
 \ee
  in the background of the external magnetic field $\vec{B}_{\rm ext}  $ and $ \theta_{\rm QED}\neq 0$. 
 This induced electric field $ \la \vec{E}\ra_{\rm ind} $   separates the electric charges, which represents the charge separation effect. 
 Then formula (\ref{E_ind}) essentially implies that the electric field locally emerges in every location where magnetic field is present in the background of the $ \theta_{\rm QED}\neq 0$.
 
The effect of separation of charges   can be interpreted  as a generation of the electric dipole moment in such unusual background.   Indeed, for a table-top type experiments it has been argued in \cite{Cao:2017ocv}  that in the presence of the  $\theta_{\rm QED} $  the electric and magnetic dipole moments of a topologically nontrivial configuration (such as a ring or torus) are  intimately related:
\be
\label{m_e}
 \la d_{\rm ind} \ra= -   \frac{\theta_{\rm QED} \cdot  \alpha}{\pi}  \la m_{\rm ind}\ra, ~~~ \alpha\equiv \frac{e^2}{4\pi}
\ee
which    obviously resembles the Witten's  effect  \cite{Witten:1979ey} when the magnetic monopole becomes the dion with electric charge $e'= - ({e\theta_{\rm QED}  }/{2\pi})  $.

To support this interpretation  we   represent the magnetic dipole moment $\la m_{\rm ind}\ra$ as a superposition of two magnetic charges $g$ and $-g$ at distance $L_3$ apart, where $L_3$ can be viewed as  the size of the compact manifold in construction \cite{Cao:2017ocv} along the third direction\footnote{This construction should be thought as a pure mathematical one. The absence of the real magnetic monopoles in Nature cannot prevent us from such fictitious   theoretical construction.}.  As the magnetic charge $g$ is quantized, $g=\frac{2\pi}{e}$,  formula (\ref{m_e}) can be rewritten as
 \be
\label{m_e_witten}
  \la d_{\rm ind} \ra= -   \frac{ \theta_{\rm QED}e^2}{4\pi^2}  \frac{2\pi L_3}{e}=-\left(\frac{e\theta_{\rm QED} }{2\pi}\right)  L_3= e' L_3~~
\ee
  This configuration becomes  an electric dipole moment
  $ \la d_{\rm ind} \ra$ with the electric charges 
 $e'= - ({e\theta_{\rm QED}  }/{2\pi})  $ which precisely coincides with the Witten's expression for $e'= - ({e\theta_{\rm QED} }/{2\pi})$  
 in terms of the $\theta_{\rm QED} $ according to  \cite{Witten:1979ey}. This construction is justified as long as magnetic monopole size is much smaller than the size of the entire configuration $L_3$ such that the topological sectors from monopole and anti-monopole do not overlap and cannot untwist themselves. The orientation of the axis $L_3$ also plays a role as it defines the $L_1L_2$ plane with non-trivial   mapping determined by  $\pi_1[U(1)]=\mathbb{Z}$,  see below. If our arguments on justification of  this formula are correct it can be applied  to all fundamental particles including electrons, neutrons, and protons
 because the typical scale $L_3\sim m_e^{-1}\sim 10^{-11} \rm cm$, while magnetic monopole itself can be assumed to be much smaller in size. In this case  
   the expression (\ref{m_e}) derived   in terms of the path integral  in \cite{Cao:2017ocv}   assumes the form
  \be
\label{d_fundamental}
\la d_{\rm ind} \ra= -   \frac{\theta_{\rm QED} \cdot  \alpha}{\pi}  \mu, 
\ee
where $\mu$ is the magnetic moment of any configuration, including the elementary particles: $\mu_e, \mu_p, \mu_n$. As emphasized in \cite{Cao:2015uza,Cao:2017ocv}  the corresponding expression can be represented in terms of the boundary terms, which normally emerge for all topological effects. 

The observed upper limit for $d_e < 10^{-29} \rm e\cdot cm$ implies that $\theta_{\rm QED}<10^{-16}$. We do not have a good explanation of why this parameter is so small. This question is not addressed in the present work. It is very possible that a different axion field must be introduced into the theory which drives $\theta_{\rm QED}$ to zero, similar to conventional axion resolution of the strong $\cal{CP}$ problem  \cite{axion1, axion2, axion3, KSVZ1, KSVZ2, DFSZ1, DFSZ2}. 
    
    The  equation  similar to (\ref{d_fundamental}), relating the electric and magnetic dipole  moments  of the elementary particles was also derived in \cite{Hill:2015kva,Hill:2015vma} where it has been argued that for time-dependent axion background the electric dipole moment of the electron  $d_e$ will be generated\footnote{We also refer to paper \cite{Flambaum:2015ica} with criticism  of this result  and \cite{Hill:2015lpa} responding to this criticism.}, and it must be proportional to the magnetic moment   of the electron $\mu_e$ and the axion  field $\theta(t)$. The absolute value for the axion field $\theta_0\approx 3.7\cdot 10^{-19}$ was 
    fixed by assuming the axions saturate the dark matter density today. While the relation (\ref{d_fundamental}) and the one derived in 
    \cite{Hill:2015kva,Hill:2015vma} look identically the same (in the static limit $m_a\rightarrow 0$ and proper normalization) the starting points are dramatically different: we begin with canonically defined  fundamental unknown constant $\theta_{\rm QED} \neq 0$
    while  computations of \cite{Hill:2015kva,Hill:2015vma} are based on assumption of time dependent axion fluctuating field saturating the DM density today, which obviously implies a different normalization for $\theta$.     Still, both expressions identically coincide in the static $m_a\rightarrow 0$ limit.  
    
    The identical expressions with precisely the same coefficients (for time dependent \cite{Hill:2015kva,Hill:2015vma} and time independent  (\ref{d_fundamental}) formulae) in  static limit $m_a\rightarrow 0$ relating the electric dipole and magnetic dipole moments   strongly   suggest that the time dependent expression \cite{Hill:2015kva,Hill:2015vma}  can be smoothly extrapolated to (\ref{d_fundamental}) with  constant $\theta_{\rm QED} $. This limiting procedure   can be viewed as a slow adiabatic process when  $\dot{\theta} \propto m_a\rightarrow 0$ and the $\theta$ becomes the time-independent parameter, $\theta\rightarrow \theta_{\rm QED}$ when the same normalization is implemented\footnote{A different approach on computation of the time dependent dipole moment due to the fluctuating $\theta$ parameter was developed recently in \cite{DiLuzio:2023ifp}.  The corresponding expression given in \cite{DiLuzio:2023ifp} approaches a finite non-vanishing constant value
      if one takes   the consecutive limits  $t\rightarrow \infty$ and after that  the static limit $m_a\rightarrow 0$ by    representing   $e/(2m)=\mu$ in terms of the magnetic moment of a fermion.    In this form    it  strongly resembles the expression derived in \cite{Hill:2015kva,Hill:2015vma}.}. 
    
We want to present one more argument suggesting that the constant $\theta_{\rm QED} $ may produce physical effects including the generating of the electric dipole moment.  Indeed the $S_{\theta}$   term in QED  in the background of the uniform static magnetic field along $z$ direction can be rewritten as follows 
 \be
\label{theta1}
S_{\theta}&\propto &  {\theta_{\rm QED} } e^2\int \dd^4 x ~\vec{E}\cdot\vec{B} =2\pi  \kappa ~\theta_{\rm QED}  \cdot \left[e \int dz dt E_z \right]. 
 \nonumber\\
&& {\rm where} ~~2\pi \kappa\equiv \left[e \int d^2x_{\perp} B_z\right] 
 \ee
 The expression on the right hand side is still a total divergence, and does not change the equation of motion. In fact, the expression in the brackets is identically the same as the $\theta$ term in 2d Schwinger model,  where it is known to be a physical parameter of the system as a result of nontrivial 
  mapping $\pi_1[U(1)]=\mathbb{Z}$, see e.g. \cite{Cao:2013na} for a short overview of the $\theta$ term in 2d Schwinger model in the given context\footnote{\label{Schwinger}In this   exactly solvable 2d Schwinger model one can explicitly see why the gauge cannot be  uniquely  fixed, and, as the consequence of this ambiguity,    the $\theta$ becomes observable  parameter of the system. The same 2d Schwinger model also teaches us how this physics can be  formulated in terms of  the so-called Kogut-Susskind ghost  \cite{Kogut:1974kt}  which is the direct analog  of the  Veneziano ghost in 4d QCD.}.
  
   The expression (\ref{theta1})
   shows once again that $\theta_{\rm QED} $ parameter in 4d Maxwell  theory  becomes  the physical parameter of the system in the background of the magnetic field\footnote{The parameter $\kappa$ which classifies our states   is arbitrary real  number. It measures the magnetic physical flux, which   not necessary assumes the  integer values.}.
    In such circumstances the electric field  will be induced along the magnetic field  in the region of  space where the magnetic field is present according to (\ref{E_ind}). 
        This relation explains why the electric dipole moment of any configuration becomes related to the magnetic dipole moment of the same configuration as equation (\ref{d_fundamental}) states. 
    
The topological arguments for special case (\ref{theta1}) when the external magnetic field is present in the system suggest that the corresponding configurations cannot  ``unwind" as the uniform static magnetic field $B_z$    enforces the system to become effectively two-dimensional, when the $\theta_{\rm QED} $ parameter is obviously a physical parameter,  similar to analogous analysis  in the well-known  2d Schwinger  model, see footnote \ref{Schwinger}.

  The practical implication of this claim is that  there are some $\theta_{\rm QED} $-dependent contributions to  the dipole moments of the particles.    
   While the $\theta_{\rm QED}$ does not produce any physically measurable effects for QED with trivial topology, or in vacuum,  we expect  that in many cases as discussed in    \cite{Cao:2017ocv}   and in present work the physics becomes sensitive to the $\theta_{\rm QED}$  which is normally ``undetectable'' in a typical scattering experiment based on perturbative analysis of QED. We want to list below several $\cal{CP}$ odd correlations which will be generated in the presence of  $\theta_{\rm QED} $, and which could be experimentally studied by a variety of instruments. 
   
   The generation of the induced electric field (\ref{E_ind}) unambiguously implies that the following $\cal{CP}$ odd correlation will be generated
   \be
 \label{EB}
 \la \vec{B}_{\rm ext}  \cdot\vec{E}\ra  = -\frac{\alpha\theta_{\rm QED}}{\pi}\vec{B}^2_{\rm ext}. 
 \ee
 Another $\cal{CP}$ odd correlation which can be also studied is as follows:
   \be
 \label{E_mu}
 \la \sum_i\vec{\mu}_{i}  \cdot\vec{E}\ra  = -\frac{\alpha\theta_{\rm QED}}{\pi}\sum_i\vec{B}_{\rm ext}\cdot\vec{\mu}_{i}, 
 \ee
where one should  average  over entire ensemble of particles with magnetic moments $\vec{\mu}_{i}$, which are present in the region
of a non-vanishing magnetic field $\vec{B}_{\rm ext}$. The induced electric field (\ref{E_ind}) will coherently accelerate the charged particles along 
$\vec{B}_{\rm ext}$ direction such that particles will assume on average non-vanishing  momentum  $\vec{p}_i$ along $\vec{B}_{\rm ext}$. As a result of this  coherent behaviour   the following  $\cal{CP}$ odd correlation for entire ensemble of particles is expected to occur
  \be
 \label{p_mu}
 \la \sum_i\vec{\mu}_{i}  \cdot\vec{p}_i\ra  \propto \frac{\alpha\theta_{\rm QED}}{\pi}\sum_i\vec{B}_{\rm ext}\cdot\vec{\mu}_{i}.
 \ee
 
 One should add that the dual picture when the external magnetic field $\vec{B}_{\rm ext}$ is replaced by external electric field $\vec{E}_{\rm ext}$ also holds. For example, instead of  (\ref{E_ind}) the magnetic field will be induced in the presence of the strong external electric field $\vec{E}_{\rm ext}$, as e.g. in the proposal \cite{Alexander:2022rmq}  to measure the proton EDM when the $\vec{E}_{\rm ext}$ is directed along the radial component, 
 \be
 \label{B_ind}
 \la \vec{B}\ra_{\rm ind} = \frac{\alpha\theta_{\rm QED}}{\pi}\vec{E}_{\rm ext},
 \ee
 such that the correlation similar to (\ref{EB}) will be also  generated 
   \be
 \label{BE}
 \la \vec{B}  \cdot\vec{E}_{\rm ext}\ra  = \frac{\alpha\theta_{\rm QED}}{\pi}\vec{E}^2_{\rm ext}. 
 \ee

            \section{Conclusion and Future Directions}\label{conclusion}
   The topic of the present notes on the dipole moments of the particles  and antiparticles   in the presence of the    $\theta_{\rm QED}$ is largely motivated by the recent experimental advances   in the  field, see \cite{Alarcon:2022ero,Alexander:2022rmq}. There are many other $\cal{CP}$ odd phenomena which accompany   
   the generation of the dipole moments.  All the relations discussed in the present notes, including (\ref{d_fundamental}) or (\ref{EB})  are topological in nature and related to impossibility to uniquely 
   describe the gauge fields over entire system,   
    as overviewed in the Introduction. 
   
    Essentially the main claim is that the $\theta_{\rm QED}$ 
     should be treated as a new fundamental parameter of the theory  when the system is formulated on a topologically nontrivial manifold, and in particular, in the background of a magnetic field which enforces a non-trivial topology, as argued in this work. 
     
     I believe that the very non-trivial  relations such as  (\ref{d_fundamental}) or (\ref{EB}) which apparently emerge in the system at non-vanishing  
$\theta$ and    $\theta_{\rm QED}$ is just the tip of the iceberg of much deeper physics rooted to the topological features of the gauge theories. 

In particular, the $\theta$ dependent portion of the vacuum energy could  be the source of the Dark Energy today (at $\theta=0$) in the de Sitter expanding space as argued in 
            \cite{Zhitnitsky:2013pna,Zhitnitsky:2015dia}. Furthermore, these highly non-trivial topological phenomena in strongly coupled  gauge theories can be tested in the QED  tabletop experiments where the very same gauge configurations which lead to the relation similar to (\ref{d_fundamental}) or (\ref{EB}) may generate an additional Casimir Forces, as well as many other effects as  discussed in \cite{Cao:2013na,Cao:2015uza,Yao:2016bps}.  What is even more important  is that  many of these effects in axion electrodynamics can be in  principle measured, see 
\cite{Tobar:2018arx,Cantatore:2018cwk,Tobar:2020kmz,Tobar:2021jyg, Tobar:2021ofd} with specific suggestions and proposals. 
I finish on this optimistic note.
                                
 \section*{Acknowledgements} 
 These notes appeared as a result of discussions with Dima Budker and Yannis Semertzidis during  the conference
 ``Axions across boundaries between Particle Physics, Astrophysics, Cosmology and forefront Detection Technologies" which took place at the Galileo Galilei Institute  in Florence, June 2023.  
 I am thankful to them  for their insisting to   write some notes on the dipole moments of the particles and their relations to the fundamental parameters of the theory, the $\theta$ and  the $\theta_{\rm QED}$. I am also thankful to participants of the  Munich Institute for Astro-, Particle and BioPhysics (MIAPbP)   workshop on ``quantum sensors and new physics" for their questions during my presentation. Specifically, I am thankful to Yevgeny Stadnik for the long discussions 
on  topics related to refs \cite{Hill:2015kva,Hill:2015vma,Flambaum:2015ica,Hill:2015lpa,DiLuzio:2023ifp}. These notes had been largely completed during the  MIAPbP  workshop in August 2023.  Therefore, I am thankful to the  MIAPbP for the organization of this workshop. The MIAPbP is funded by the Deutsche Forschungsgemeinschaft (DFG, German Research Foundation) under Germany's Excellence Strategy- EXC-2094 -390783311. This research was supported in part by the Natural Sciences and Engineering
Research Council of Canada.

  \bibliography{Axion-theta}

\end{document}